\documentclass[runningheads]{llncs}

\usepackage{booktabs}
\usepackage{xcolor}

\usepackage{placeins}
\usepackage{hyperref}

\newcommand{\coma}[1]{{\color{red}[AV: #1]}}

\usepackage{xspace}
\newcommand{\mv}{MultiVeStA\xspace}
\newcommand{\mq}{MultiQuaTEx\xspace}
\newcommand{\nl}{NetLogo\xspace}
\newcommand{\autow}{\texttt{autoWarmup}\xspace}
\newcommand{\autobm}{\texttt{autoBM}\xspace}
\newcommand{\autord}{\texttt{autoRD}\xspace}
\newcommand{\autoW}{\autow}
\newcommand{\autoBM}{\autobm}
\newcommand{\autoRD}{\autord}
\newcommand{\manualbm}{\texttt{manualBM}\xspace}
\newcommand{\manualrd}{\texttt{manualRD}\xspace}
\newcommand{\transient}{\texttt{autoIR}\xspace}

\usepackage{listings}
\newcommand{\lline}[1]{Line~\ref{#1}} 
\newcommand{\llines}[2]{Lines~\ref{#1}-\ref{#2}} 

\usepackage{subfig}
\usepackage{graphicx}
 
\title{Statistical Model Checking of NetLogo Models}

\author{Marco Pangallo\inst{1}\and Daniele Giachini\inst{2} \and Andrea Vandin\inst{2,3}}

\institute{CENTAI, Turin, Italy. \and
	Institute of Economics and L'EMbeDS, Sant'Anna School of Advanced Studies, Pisa, Italy.
	\and
	DTU Technical University of Denmark, Lyngby, Denmark.
	}

\usepackage{natbib}
\setcitestyle{authoryear,round,aysep={}}
	
\usepackage{amsmath}

\begin{document}
\maketitle 



\begin{abstract}
Agent-based models (ABMs) are gaining increasing traction in several domains, due to their ability to represent complex systems that are not easily expressible with classical mathematical models.
This expressivity and richness come at a cost: ABMs can typically be analyzed only through simulation, making their analysis challenging. Specifically, when studying the output of ABMs, the analyst is often confronted with practical questions such as: (i) how many independent replications should be run? (ii) how many initial time steps should be discarded as a warm-up? (iii) after the warm-up, how long should the model run? (iv) what are the \emph{right} parameter values? Analysts usually resort to rules of thumb and experimentation, which lack statistical rigor. This is mainly because addressing these points takes time, and analysts prefer to spend their limited time improving the model. 
In this paper, we propose a methodology, drawing on the field of Statistical Model Checking, to automate the process and provide guarantees of statistical rigor for ABMs written in NetLogo, one of the most popular ABM platforms. We discuss \mv, a tool that dramatically reduces the time and human intervention needed to run statistically rigorous checks on ABM outputs, and introduce its integration with \nl.
Using two ABMs from the \nl library, we showcase \mv's analysis capabilities for \nl ABMs, as well as 
a novel application to statistically rigorous calibration. Our tool-chain makes it immediate to perform statistical checks with \nl models, promoting more rigorous and reliable analyses of ABM outputs.
\end{abstract}

\begin{keywords}
NetLogo, MultiVeStA, Transient analysis, Calibration, Warmup estimation, Steady-state analysis
\end{keywords}






\section{Introduction}


You have just finished coding your ABM and are starting to analyze its output. But critical questions arise: \textit{How many simulation runs are enough? How many steps should be discarded as warm-up? Once warm-up is discarded, how long should the simulation continue? 
}
 These questions are essential to obtain statistically valid results and ensure that outcomes are not driven by a single low-probability realization. Yet, in practice, there's often no clear answer. So you settle on ``reasonable'' numbers---say, 10 simulations, 50 warm-up steps, and 100 steps for steady-state analysis---largely driven by computational constraints. But you are left wondering: \textit{Was that sufficient? Or did I overdo it, wasting computational resources?}

The main reason why more principled methods are not used is that they require substantial effort to implement correctly. Statistically rigorous rules are time-consuming to design and code, and integrating them within the ABM increases the risk of bugs and slows down model development. Modelers often prefer to spend their limited time enhancing the realism of the model rather than coming up with bullet-proof statistical guarantees.

In this paper, we present an automated solution to these challenges. Our approach is simple enough that the modeler can focus solely on the ABM logic, while outsourcing the statistical analysis to a dedicated framework. The modeler only needs to define (i) the ABM step function and how to reset the random seed, and (ii) which model variables should guide the determination of simulation runs, warm-up length, and steady-state duration. Once this minimal information is provided, our framework runs the smallest number of simulations needed to produce statistically valid results, efficiently parallelizing independent runs. It then returns a complete analysis to the user.

Our method is grounded in the well-established field of Statistical Model Checking (SMC), which combines statistical inference and computer science to deliver optimal solutions. SMC is based on two core principles. First, the analyst should define the \textit{properties of interest}, and the tool should automatically run the minimal number of simulations needed to estimate those properties within a given confidence level. Second, the statistical analysis should be decoupled from the model implementation. This means the SMC framework can be written in a separate programming language, with its own syntax, communicating externally with the ABM.

The specific contribution of this paper is to integrate an SMC tool, \mv{}, with the most widely used ABM platform, \nl. At the same time, we demonstrate the capabilities of the tool-chain through a set of illustrative examples and extend standard SMC to issues related to parameter calibration. Our aim is to raise awareness among agent-based modelers of how SMC can facilitate statistically rigorous output analysis. We walk the reader through concrete applications, similarly to \cite{thiele2014facilitating,JSSv058i02}, and show how SMC can be used to build a more nuanced view of what parameter combinations best explain real data.

The first SMC method we present is \textit{transient analysis}, which focuses on obtaining statistically valid results at specific time points. The SMC tool determines the number of simulation runs required so that the mean value at a given step lies within a user-specified confidence interval and significance level. Some parts of the simulation may be more volatile than others: in such cases, the SMC automatically allocates more simulations to the uncertain periods, reducing computational burden where possible. Most importantly, these steps are fully automated.

To illustrate transient analysis, we consider the ``Artificial Anasazi'' model introduced by \cite{dean2000understanding, axtell2002population} and further explored by \cite{janssen2009understanding}. We use the \nl implementation from \cite{stonedahl2010netlogo}. The model simulates population dynamics in northeastern Arizona between 800--1350 AD. The key determinant of population dynamics is the potential agricultural production, fluctuating with droughts, and pinned down from real paleoenvironmental data. Our transient analysis reveals that matching simulated and real household data can be inaccurate during periods of rapid growth---especially when only a single ``representative'' simulation is considered, as in \cite{axtell2002population}, or even when averaging over 15 runs as in \cite{janssen2009understanding}. In contrast, our method runs up to 600 simulations during uncertain periods to ensure statistically reliable results, while saving resources by running only 20 simulations during stable phases.

Building on this, we introduce a statistically rigorous approach to calibration that goes beyond the traditional search for a single optimal parameter set. Replicating the 5-parameter calibration exercise of \cite{janssen2009understanding}, we obtain a parameterization close to, but not identical with, their reported optimum. Crucially, rather than treating this outcome as a unique solution, our framework evaluates whether alternative parameter combinations perform significantly better or worse. By formally computing $p$-values for pairwise comparisons, we show that 14 parameter combinations cannot be statistically distinguished in terms of fit. This advances standard calibration practices by shifting the focus from point estimates to confidence sets of plausible parameterizations, in line with the \emph{model confidence set} approach of \cite{hansen2011model} and its application to ABMs by \cite{seri2021model}. In doing so, our method provides an automated and statistically guaranteed procedure for calibrating \nl models. Importantly, this does not only provide a technical improvement, but allows one to achieve stronger conclusions on the behavior of the model. For instance,  our procedure lets us confirm with robust evidence that harvest-related parameters dominate demographic ones in explaining historical population dynamics.

We also address two additional challenges: (i) determining the length of the warm-up period, during which simulation dynamics are dominated by random initial conditions; and (ii) estimating how many steps are required in the pseudo-steady state (a phase of fluctuating but stationary behavior) to compute reliable long-run statistics. Warm-up detection is performed by dividing the simulation into batches and testing whether variables across batches follow a stable distribution (e.g., normal). The simulation continues until this condition is met. Then, steady-state statistics are computed from the post-warm-up data, with the number of runs determined by user-defined significance level and confidence intervals.

To demonstrate these features, we use the ``Alpha Birds'' model from \cite{railsback2012agent}.\footnote{This ABM was also used by \cite{thiele2014facilitating} to illustrate their RNetLogo package that integrated \nl and the R programming language.} This ABM simulates territorial bird populations with reproductive suppression driven by dominant alpha birds. In the original application, 10 simulations were run for 20 years, after discarding 2 years as warm-up. Our framework confirms that this approach was adequate for certain outputs, but not for all. In particular, the model features a sharp transition: when a survival probability parameter falls below a threshold, extinction becomes likely. Yet extinction can take longer than 2 years to occur. As a result, the original warm-up period was insufficient, and the analysis missed a key discontinuity in model behavior.

The rest of the paper is organized as follows. Section~\ref{sec:prelim} introduces SMC, \mv, and \nl, and explains their integration and query structure. Section~\ref{sec:anasazi} demonstrates transient analysis using the Artificial Anasazi ABM, including its application to calibration. Section~\ref{sec:alphabirds} illustrates warm-up and steady-state analysis with the Alpha Birds model. Section~\ref{sec:discussion} concludes.




\section{Preliminaries}\label{sec:prelim}
\subsection{Agent-Based Modeling and NetLogo}\label{sec:nl}
NetLogo is a popular platform for agent-based modeling (ABM), a computational approach to simulate the actions and interactions of autonomous agents to assess their effects on the system as a whole. It uses its own language, called NetLogo, too, which is designed to be simple and accessible, making it particularly useful for both beginners and experts in the field. NetLogo excels at visualizing and exploring complex systems, allowing users to observe emergent behaviors resulting from individual agents' rules. Its flexibility and ease of use make it ideal for developing, running, and experimenting with agent-based models in a wide range of fields.

NetLogo is widely regarded as the most common platform for agent-based modeling. Its extensive Models Library provides a rich repository of pre-built models from various disciplines, including biology, economics, sociology, and environmental science. Researchers, educators, and students use NetLogo for academic studies, teaching complex systems concepts, and simulating real-world scenarios. The platform's interactive interface, combined with its ability to run on standard computers, has contributed to its status as a go-to tool in the ABM community.

\subsection{Statistical Model Checking and MultiVeStA}\label{sec:mv}
Statistical Model Checking (SMC) is a family of automated simulation-based analysis techniques popular in Computer Science (see, e.g.,~\citep{Agha18,LLTYSG19}). 
All SMC proposals base on the idea of automatically performing a \emph{sufficient} but \emph{minimal} number of stochastic simulations (and simulation steps) of a model to obtain \emph{statistically reliable estimations} of given model \emph{properties of interest}. 
Furthermore, following the so-called principle of \emph{separation of concerns}, properties to be analysed are not encoded inside the model itself (i.e., to compute intermediate quantities, or create/modify CSV files), but are expressed in an external query language, a \emph{property specification language}. This is similar to how, in databases, queries of interest are expressed in SQL without requiring to modify the database definition for any different study. This tends to keep model specifications cleaner and favor replicability of experiments.
%
In some sense, statistical model checkers can be seen as libraries of analysis routines that can be run with ``one-click'' and that guarantee that the obtained results are statistically reliable. This makes it easier to engineer well (part of) the analyses and validation tasks.
So-called black-box SMC (e.g.,~\citet{sen2004statistical,younes2005probabilistic,mvjedc2022}) do not make any assumption on the nature of the stochastic process underlying the studied model. They only assume that the model can be (stochastically/probabilistically) simulated. For example,  \mv{}~\citep{mvjedc2022,sebastio2013multivesta,DBLP:conf/ifm/GilmoreRV17} is a black-box SMC tool that can be integrated with existing simulators to enrich them with automated statistical analysis techniques. \mv{} has been 
recently redesigned and extended to target agent-based models (ABM) from the social sciences~\citep{mvjedc2022}.

\citet{mvjedc2022} have shown that SMC can help in  the analysis of ABMs from the social sciences, promoting the automation of statistically-reliable analysis, and avoiding errors in the results  arising from wrong analysis designs. 
For example, \cite{Secchi2017} performed a study on 55 ABMs published between 2010 and 2013 in high-quality venues from the management and organizational science. The authors demonstrated how, in most cases, simulation exercises did not offer acceptable statistical quality, rising doubt on the results and their implications. Similarly, a poor handling of initial warm-up behaviours can distort results and the interpretation of steady-state behaviours \citep[at equilibrium, see, e.g.,][]{gal_aacute_n2005}. \citet{mvjedc2022} show that SMC, and in particular \mv{}, can solve these issues automatically, demonstrating it on two ABMs from the literature. 
%
%
First, they show how to automate analyses performed on a large macroeconomic ABM~\citep{caiani2019does}, obtaining higher statistical reliability than classic analysis setups used by the ABM  community, and scaling down the analysis time from days to hours. This focused on \emph{transient analyses}, i.e., analyses done on specific time points in the simulation (each of the first 400 simulation steps, each corresponding to three months). After this, the authors moved to \emph{steady-state analyses}, i.e., analyses done after the system stabilizes in a steady state. This is a much more complex type of analysis. The authors show how to identify and solve qualitative and quantitative analysis errors performed on an ABM of a financial market~\citep{kets2014}, focusing on market selection and price at steady-state.

\subsection{Integrating NetLogo and MultiVeStA}\label{sec:nl+mv}
\mv{} has been created to allow for easy integration with new simulators. For example, it has been integrated with Java-, C-, R-, Matlab- and Python-based simulators~\footnote{\url{https://github.com/andrea-vandin/MultiVeStA/wiki}}.
In this paper, we present the integration of \mv and \nl. 
In order to integrate a new simulator with \mv, one has to implement a piece of software, an \emph{adaptor}, between \mv and the simulator. Such adaptor only needs to expose to \mv{} 3 basic actions present in any existing simulator: 
\begin{enumerate}
	\item \texttt{reset(seed)}: to reset the simulator to its ``initial state'', and update the random seed used to generate pseudo-random numbers. This is necessary to reset the model before performing a new simulation, and to allow \mv to take care of random-seed generation (with the possibility of controlling the generated seeds for replicability);
	\item \texttt{next}: to perform one step of simulation, where one \emph{step} may correspond, e.g., to one time unit (tick) in \nl , or more, depending on modeling needs;
	\item  \texttt{eval(obs)}: to evaluate an observation (\texttt{obs}) in the current simulation state, where \texttt{obs} can be any feature of the aggregate model, of groups of agents, or even of individual agents. 
	In particular, in this paper an observation \texttt{obs} can be any arithmetic expression that can be queried in the \nl console.
\end{enumerate}

%

More in detail, the integration of \mv and \nl has been made possible using the \nl{} Java APIs~\footnote{\url{https://ccl.northwestern.edu/netlogo/2.1/docs/controlling.html}}. These APIs allow \mv{} to access \nl programmatically, enabling the support for \texttt{reset}, \texttt{next} and \texttt{eval}. Any \nl{} model is now natively supported by the tool-chain presented in this paper, without requiring any modification. 
This approach is followed, e.g., also by RNetLogo~\cite{thiele2014facilitating,JSSv058i02}, which uses such APIs to access programmatically \nl models using the R programming language.

\subsection{\mv{} query language and supported analysis}\label{sec:queries}
\begin{figure}[t]
	\centering
	\subfloat[Transient analysis]{\includegraphics[scale=0.5]{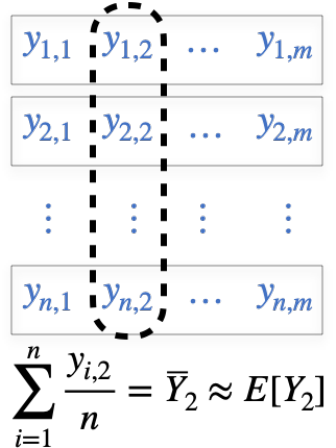}}
	\hfill
	\subfloat[Steady-state analysis by \emph{Replication and Deletion} (RD)]{\includegraphics[scale=0.5]{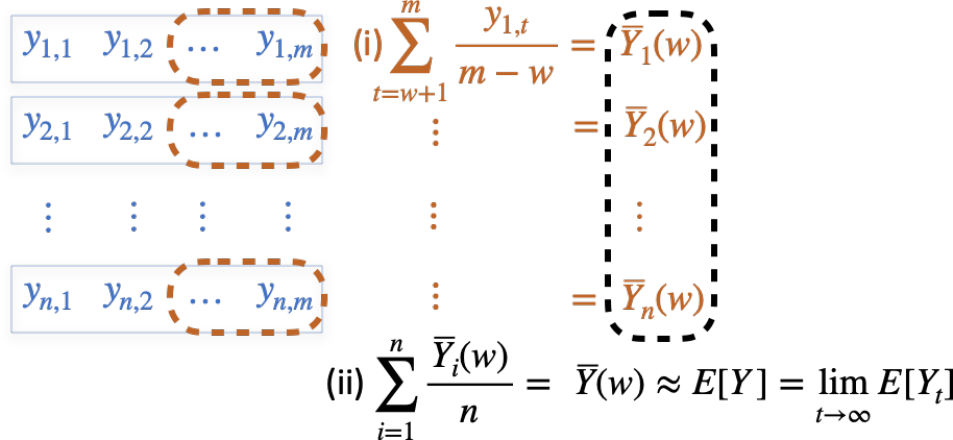}}
	\\
	\subfloat[Steady-state analysis by \emph{Batch Means} (BM)]{\includegraphics[scale=0.5]{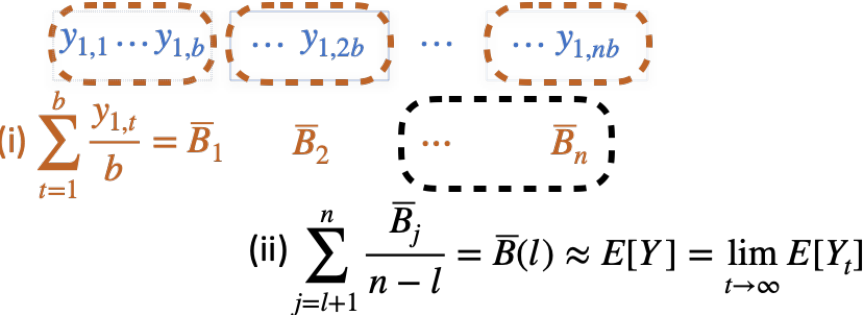}}
	\caption{Transient and steady-state analysis in \mv{}. Panels (a) and (b) use \emph{n} simulations of \emph{m} steps each. Panel (c) uses one long simulation. Adapted from~\cite{mvjedc2022}.  
    }
	\label{fig:nmsimsAnalysis}
\end{figure}

In this section we informally review from an user perspective the analysis capabilities offered by \mv{}, and refer to~\cite{mvjedc2022} for more details. 
The analysis of ABMs often  builds on stochastic simulations, based on Monte Carlo methods
to estimate model characteristics \citep[see, e.g.,][]{richiardi2006common,lee2015complexities,fagiolo2019validation}. 
In the majority of cases,  we can think of (the output of) an ABM as a collection of discrete-time stochastic processes $(\mathbf{Y}_t)_{t>0}$ describing the evolution over time $t$ of variables of interest (e.g., the wealth of an agent, the average population count of a category of agents, etc.).
For easiness of presentation, let us focus on the case in which $(\mathbf{Y}_t)_{t>0}$ contains only one time series of interest $(Y_t)_{t>0}$. I.e., $Y_t$, for each $t>0$, is a random variable of interest defined over the ABM. 
%
Figure~\ref{fig:nmsimsAnalysis}(a) provides a pictorial representation of $n$ independent replications of  a single variable 
$Y_t$ (one per row) each comprising $t=1,\ldots, m$ steps (one per column).
\footnote{With the term  {\em independent replications} we mean runs (or simulations) obtained each using different random seeds and resetting the simulator status reset to an initial configuration.}
The outcome of a simulation $i$ can be represented as  $y_{i,1},\ldots,y_{i,m}$, a sequence of observations (or realizations) 
 of length $m$.
%
Clearly, the observations within the same column $t$ are independent and identically distributed (IID), while those in the same row $i$ are not. 

As mentioned in the introduction, and as discussed in detail in~\cite{mvjedc2022}, \mv{} focuses on two typical classes of properties:
\begin{itemize}
	\item \emph{Transient properties}: the expected value of a property of a model at a given 
	time $t$ (or 
	within a time range, or at the
	occurrence of a specific event);
	As depicted in Figure~\ref{fig:nmsimsAnalysis} (a), for a given point of interest $t$ this can be expressed as $E[Y_t]$. 
	\item \emph{Steady-state properties}: the expected value of a property of a model when it has reached a statistical equilibrium (i.e., at steady state). 
	As depicted in Figure~\ref{fig:nmsimsAnalysis} (b), this can be expressed as $E[Y]\!=\!\lim_{t\to\infty} E[Y_t]$.
\end{itemize}
%
Examples of transient properties are given in Section~\ref{sec:anasazi}, where we study the average number of households alive for each of the 550 time points of interest for the artificial Anasazi model~\citep{dean2000understanding,axtell2002population}. Here, a transient analysis is necessary because the model was proposed to replicate pointwise historical data, without any interest for steady states. In order to estimate such properties, \mv{} offers the method \transient discussed later. 
Instead, an example of steady-state property can be found in Section~\ref{sec:alphabirds}. Here, we consider a typical ABM in ecology \citep{railsback2012agent} developed to study long-term survival ratios of birds species. Here, steady-state analyses are necessary because the modeler is interested in steady-state properties. 
Clearly,  steady-state analyses are meaningful only ``around'' a statistical equilibrium. That is,   $\lim_{t\to\infty} E[Y_t]$ shall exist and be finite. To automatically check when this assumption holds, \cite{mvjedc2022} proposed a diagnostics. For the cases in which it holds, they proposed two steady-state methods, namely \autoRD, based on the so-called \emph{Replication and Deletion} (Figure \ref{fig:nmsimsAnalysis} (b)), and \autoBM, based on \emph{Batch Means} (Figure \ref{fig:nmsimsAnalysis} (c)).  
Both rely on an algorithm  for estimating the end of the initial warmup period of the simulator, named \autoW. This is used by \autoRD and \autoBM to discard initial steps of simulation which are affected too much by the initial conditions, washing out the initial transient phase of the model. Then, the two approaches differentiate in that \autoRD performs several \emph{short} simulations, while \autoBM performs a long simulation.

\mv offers an intuitive \emph{property specification language}, \mq. Using \mq, one can express both transient and steady-state properties. In the latter case, one can also just check for the end of the initial warmup period of the model, necessary for steady-state analysis. 

\FloatBarrier

\subsubsection{\mq for transient properties}
A \mq query for transient analysis can describe one or more random variables evaluated on a model, each evaluated using the same simulations and with their own confidence interval (CI). 
Listing~\ref{ls:examplePropertyTr} depicts a \mq{} query used in Section~\ref{sec:anasazi} to study the evolution of (i) the number of households (line 5), and (ii) the difference between the simulated number of households and the historical data (line 6), in the first 570 simulation steps (line 7).
%
Therefore, it considers 1140 random variables, meaning that each simulation will give one sample for each of the 1140 random variables. 
Following the discussion in Section~2 of \cite{mvjedc2022}, the expected value of each random variable denoted by a \mq query is estimated as the mean $\overline{Y}$ of~$n$ samples (coming from  $n$ independent replications), with $n$ large enough but minimal to guarantee that the  $(1-\alpha) \cdot 100\%$ CI centred on $\overline{Y}$ has size at most $\delta$, for given $\alpha$ and $\delta$. 

More specifically, the user has to provide \mv{} with a desired statistical significance level $\alpha$ and  interval width $\delta$. With this, \mv{} uses the method \texttt{autoIR} which automatically performs the minimum number of independent replications that let each point estimate be in a confidence interval of width at most $\delta$, with statistical confidence $1-\alpha$. In particular, \mv{} performs the minimum number of independent replications for each point. 
It does so by performing an initial number $n$ of replicas (user-specified, default 20), computes the mean value $\overline{Y}_t$ at each time-step, and calculates the confidence interval as
\begin{equation*}
    \overline{Y}_t\pm \boldsymbol{t}_{n-1,1-\frac{\alpha}{2}}\sqrt{\frac{s^2_t}{n}}\,,
\end{equation*}
where $\boldsymbol{t}_{n-1,1-{\alpha}/{2}}$ refers to the $1-{\alpha}/{2}$ quantile of a Student-t distribution with $n-1$ degrees of freedom and $s^2_t$ is the unbiased estimator of the variance. If the width of the confidence interval, $2\boldsymbol{t}_{n-1,1-{\alpha}/{2}}\sqrt{{s^2_t}/{n}}$, is smaller than $\delta$ for some $t$, then the corresponding estimates are stored, and those points are not considered anymore in future simulations. This means that, even if more simulations are run, the estimates for these points will ignore future samples. If for some $t$, instead, the interval width is larger than $\delta$, further $n$ independent replicas are performed. The mean value and confidence interval computations are iteratively computed, 
considering each time $n$ more simulations. At each iteration, more points may reach the required confidence interval width, and will therefore be ignored in future iterations. The algorithm keeps iterating by adding more independent replicas until all points have confidence intervals smaller than $\delta$. When this happens, the results are shown \citep[for further detail, see][]{mvjedc2022}. Of course, the procedure can independently estimate more variables within the same simulations. 
%

%

\lstset{caption=A transient MultiQuaTEx query, label=ls:examplePropertyTr, basicstyle=\small\ttfamily}
\begin{lstlisting}[float=t,mathescape,morekeywords={next,autoIR,parametric,eval,E,if,fi,then,else,s,eval,evalME,evalOnceME},numbers=left,belowskip=-8pt]
obsAtStep(step,obs) = if (s.eval("steps") == step)!\label{ls:if}! !\label{ls:examplePropertyTrif}!!\label{ls:rval}!
			then s.eval(obs) !\label{ls:rval2}!
			else next(obsAtStep(step,obs))!\label{ls:examplePropertyTrNext}!
		      fi ;!\label{ls:examplePropertyTrfi}!
eval autoIR(E[ obsAtStep(step,"tothouseholds") ],
	    E[ obsAtStep(step,"abs(tothouseholds - histothouseholds)") ],
	    step,0,1,570) ;!\label{ls:examplePropertyTrParametric}!
\end{lstlisting}
We now move our attention to the structure of the query. It starts with a list of \emph{parametric operators} 
that can be used in an \texttt{eval} \transient command to specify the properties to be estimated. 
\llines{ls:examplePropertyTrif}{ls:examplePropertyTrfi} of Listing~\ref{ls:examplePropertyTr} define the parametric operator \texttt{obsAtStep} having two parameters, \texttt{step}  and \texttt{obs}, respectively the step and observation of interest. 
Lines 5-7
instantiate such operator 
for each of the 1140 random variables considered, allowing to obtain 1140 corresponding samples per simulation. 
%
%
%
As shown in \llines{ls:rval}{ls:examplePropertyTrfi}, in the operator definition we may use \texttt{s} to denote the current simulation state, the function \texttt{eval} to evaluate expressions in \texttt{s} (e.g., the current number of performed \texttt{steps} and the observations \texttt{obs} specified in lines 5-6, or expressions using \nl's query languages), and the function \texttt{next} to ask the simulator to perform one step of simulation. 
%
%
More in detail,  an operator might be defined by means of: 
\begin{enumerate}
	\item conditional statements (the \texttt{if}-\texttt{then}-\texttt{else}-\texttt{fi}); 
	\item real-valued observations on the current simulation state (the \texttt{s.eval} in \lline{ls:rval} and \lline{ls:rval2}); 
	\item a \texttt{next} operator 
	%
	\item arithmetic expressions.
\end{enumerate}
%
%
Building on this, we can see that, in Listing~\ref{ls:examplePropertyTr}, we check whether we have reached the step of interest (\lline{ls:if}), in which case we return the required observation (\lline{ls:rval2}). Otherwise,   we perform a step of simulation (\lline{ls:examplePropertyTrNext}), and evaluate recursively the operator in the next simulation state. 

%

\subsubsection{\mq for steady-state properties and warmup analysis}
\mq{} also  supports  the steady-state and warmup analysis capabilities of \mv introduced in~\cite{mvjedc2022}. 
Listing~\ref{ls:examplePropertySS} provides a \emph{steady-state} \mq query used in Section~\ref{sec:alphabirds} to study at steady-state the average population of birds, and other properties. Notably, the \texttt{count}, \texttt{with}, etc shown in the query are keywords of \nl's query language which can be used (similarly to the transient case) to evaluate complex expressions on the current simulation state. 
The structure of the query is simpler than the one for transient analyses, as the progressing of simulation steps is performed implicitly in the analysis procedure. For this reason, we replace the operator \texttt{obsAtStep} with the simpler operator  \texttt{obs} which just returns the observation of interest in the current simulation step.  The query is then completed by one of the three types of supported steady-state analyses shown in  \llines{ls:ssan1}{ls:ssan3}.
\lstset{caption=A steady-state MultiQuaTEx query. Only one of the three eval commands should be used at a time., label=ls:examplePropertySS}
\begin{lstlisting}[float=t,mathescape,morekeywords={warmup,autoRD,autoBM,parametric,eval,E,if,fi,then,else,s,rval,evalME,evalOnceME},numbers=left,belowskip=-8pt]
	obs(o) = s.eval(o) ;
	eval autoWarmup(E[obs("count turtles") ],
		E[obs("count patches with [count(turtles-here with [is-alpha?]) < 2]") ];!\label{ls:ssan1}!
	eval autoBM(E[obs("count turtles") ],...) ;!\label{ls:ssan2}!
	eval autoRD(E[obs("count turtles") ],...) ;!\label{ls:ssan3}!
\end{lstlisting}
In particular, a steady-state query is composed of two parts:
\begin{enumerate}
\item 
A list of \texttt{next}-free operators,  
\item 
 one of the three \texttt{eval} commands in Listing~\ref{ls:examplePropertySS}, provided with a list of operators to consider. 
\end{enumerate}
%
In particular, \autoW performs the warmup estimation procedure from~\cite{mvjedc2022} for each of the listed properties. This is because every random variable on a process might have a different warmup period (see the discussion in~\cite{mvjedc2022}). 
Further to estimating the warmup period for each property, \autoBM and \autord  run the batch means or replication and deletion procedures from~\cite{mvjedc2022} to estimate each property in the steady state, discarding the corresponding warmup periods.  The estimations performed by \autoBM and \autord are equipped with $(\alpha,\delta)$ confidence intervals as for transient properties. 
%

\mq supports two further \texttt{eval} commands: \manualbm and \manualrd. These behave similarly to \autobm and \autord, but take the warmup length as an input, rather than estimating it. 
This is useful in case this information is already known. In Section~\ref{sec:alphabirds} we use them to replicate analyses from the literature.

\FloatBarrier

\section{Transient Analysis and Calibration application: Artificial Anasazi model}\label{sec:anasazi}

In this section,, we apply our transient analysis techniques to the ``Artificial Anasazi'' model originally proposed by \cite{dean2000understanding} and \cite{axtell2002population}, and further explored by \cite{janssen2009understanding}. For all our applications, we use the \nl{} implementation provided by \cite{stonedahl2010netlogo}.\footnote{The model is available at \url{https://ccl.northwestern.edu/netlogo/models/ArtificialAnasazi}.}
The model is meant to reproduce the dynamics of the Kayenta Anasazi population dwelling in the Long House Valley in the Black Mesa area of northeastern Arizona (U.S.) between 800 and 1300 AD. The ecological landscape of the Long House Valley -- in particular, the annual fluctuations in potential agricultural production -- is closely reconstructed using paleoenvironmental research data. Such a landscape is populated by artificial agents representing households that are initialized on nutritional requirements and demographic characteristics from ethnographic studies.
Each household has agent-specific attributes (e.g., age, size, composition, amount of maize consumed, maximum amount of maize that can be stored, use of total potential maize yield) and interacts with others according to behavioral rules. More specifically, the rules determine where the household should locate its maize plantation and dwelling. In each period $t=800,\ldots,1350$, corresponding to a calendar year, every agent performs agricultural activities, harvests, consumes, and can change its plots and/or dwelling location  based on whether its nutritional needs have been met or not. Moreover, the population evolves over time, since, depending on their nutritional success, households can reproduce or disappear. 
At the end of each period, the total number of households and their location are recorded, such that the artificial population dynamics can be compared with the real dynamics reconstructed using archaeological and environmental data. 

In this setting, \cite{axtell2002population} show how, after calibrating its parameters, the model can closely reproduce key spatial and demographic features of the dynamics of Anasazi population. \cite{janssen2009understanding} deepens the previous analysis  focusing on the total number of households. The author considers five key parameters (death age, end of fertility age, fission probability, harvest adjustment, harvest variance) and shows that accurate calibration only partially contributes to the good fit of the model. Instead, the exogenously given carrying capacity (that is, the number of locations that have a yield equal or higher than the nutrition needs) is the fundamental driver of the success of the model in matching the data. Hence, harvest-related parameters appear to be more important than the demographic ones in explaining the model performance.

In establishing the fitness of each combination of parameter values, both \cite{axtell2002population} and \cite{janssen2009understanding} use the same methodology. They perform 15 independent replications (simulations) with different random seeds. For each independent replication, they compute the average distance (over time) of the model-generated number of households from the empirical time series, using either the $L^1$ or the $L^2$ norm. Then, the smallest value of average distance, among the 15 computed, is assigned to the parameter combination as its fitness measure. The best parameter combination is selected as the one minimizing the fitness measure over the set of combinations considered. In showing the results, \cite{axtell2002population} argue that:
%
 \[\text{
 ``\emph{the average run, produced by averaging over 15 distinct runs, looks quite similar}''  
}\]
 %
 (to the one they show relative to the best parameter combination). 
 At the same time, \cite{janssen2009understanding}, simulating the model 100 times using the best combination of \cite{axtell2002population}, notices 
 %
 	\[\text{
  	``\emph{some variation in the results}''.
  }\]
 %
Indeed, the outcomes of the Artificial Anasazi model, similarly to those of the vast majority of agent-based models, can be profoundly affected by the intrinsic stochasticity deriving from the sequence of random draws characterizing its functioning. As a consequence, on the one hand, any measure of fitness that is attached to a parameter combination on the basis of a given number of independent replications is itself a random variable. Hence, it suffers from estimation errors that need to be quantified and assessed when making decisions such as choosing the best parameter combination. At the same time, any assessment of the concordance between real and artificial data should take into account the variability that the simulation exercise entails. 
Neither \cite{axtell2002population} nor \cite{janssen2009understanding} deal with these potential problems in their calibration exercises and concordance assessments. 
Hence, in what follows we perform two computational exercises that highlight the potential of the transient analysis capabilities of \mv{} in both performing calibration tasks with statistical guarantees and in reliably assessing the performance of a given parameterization of the model. For exposition purposes, we shall consider first the concordance assessment using the best combination individuated by \cite{janssen2009understanding} and, then, we shall perform the calibration exercise with statistical guarantees. In this way, we can first present the main features of transient analysis and, then, explain its application to calibration of computational models.

\subsection{Transient analysis of the best parameter combination by \cite{janssen2009understanding}}

\begin{figure}[!t]
\centering
\includegraphics[width=0.5\textwidth]{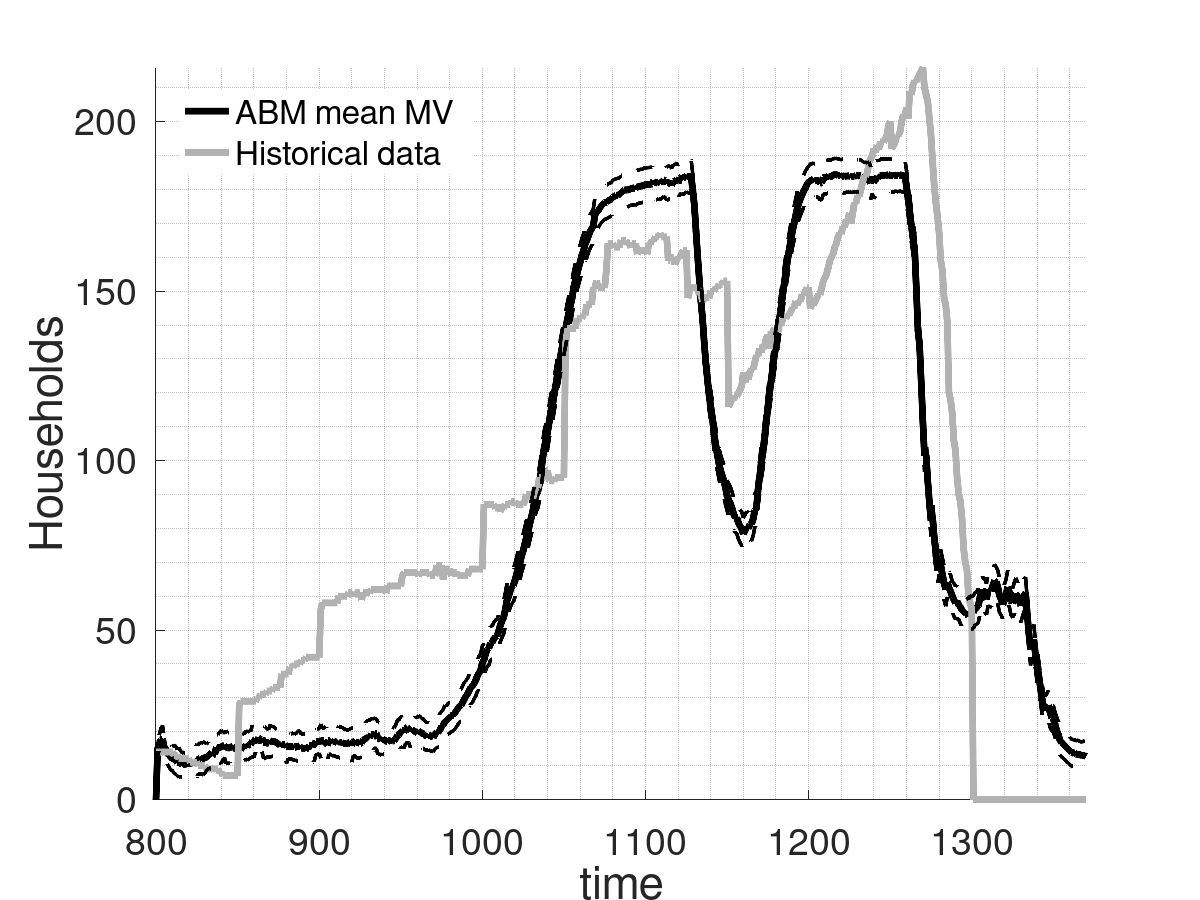}\includegraphics[width=0.5\textwidth]{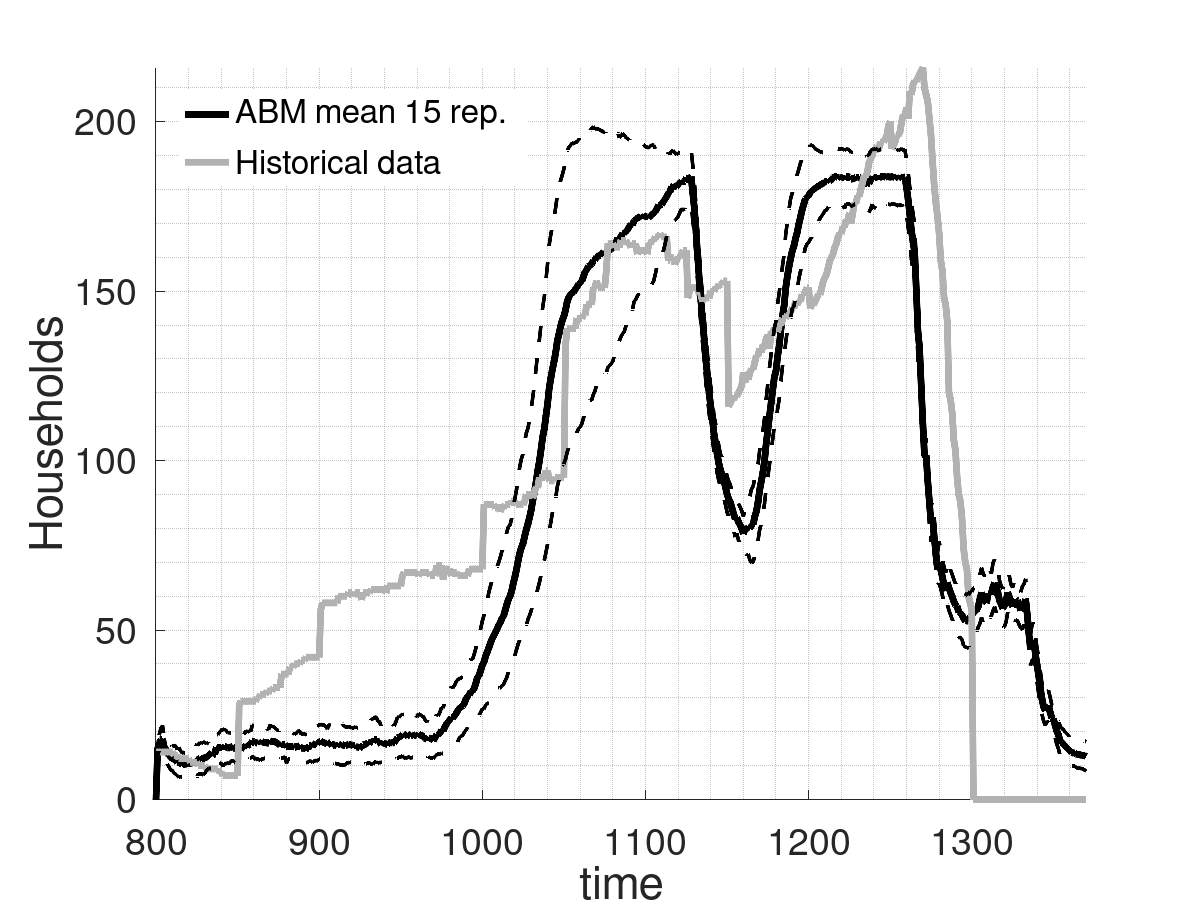}\\
\includegraphics[width=0.5\textwidth]{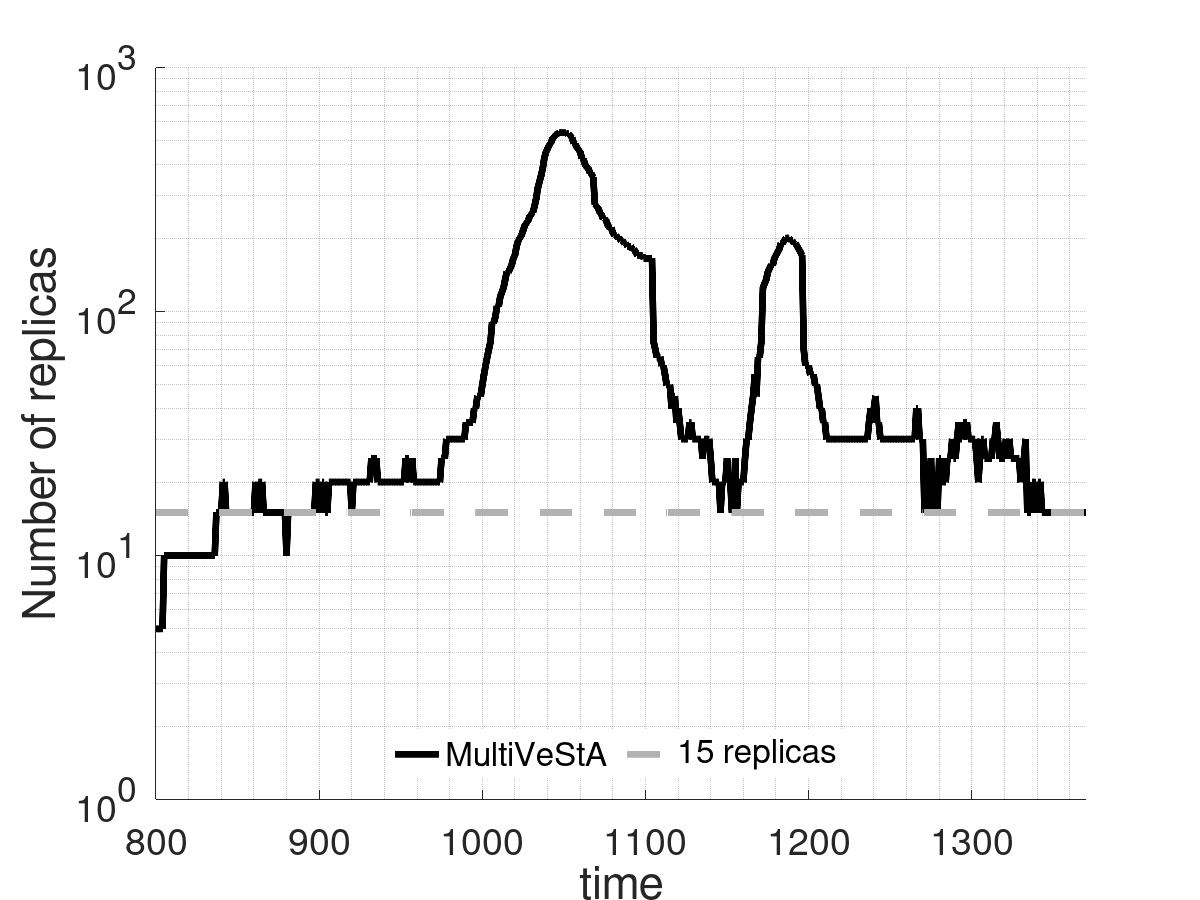}
\caption{Analysis of the total number of households in the Artificial Anasazi model under the best combination of parameters by \cite{janssen2009understanding}: death age $38$, end of fertility age $34$, fission probability $0.155$, harvest adjustment $0.56$, harvest variance $0.4$. \textbf{Top left}: \mv{}'s transient analysis with significance level $\alpha=0.01$ and width of the confidence interval $\delta=10$. In black, we provide the estimation (with confidence intervals dashed), while in grey the real data. \textbf{Top right}: Same as top-left panel, but we impose the use of exactly 15 independent replications (as in \cite{janssen2009understanding}). \textbf{Bottom}: number of replicas used by 
\mv{} for the transient analysis in the top-left panel.}
\label{fig:anasazi_rep}
\end{figure}

In our first exercise we consider one variable of interest, the total number of households, estimated for each $t=800,\ldots,1350$.
In Figure \ref{fig:anasazi_rep}, we present the outcomes of our analysis under the best combination of parameters by \cite{janssen2009understanding}. 
In the top-left panel we present the results of the transient analysis obtained setting significance level $\alpha=0.01$ and width of the confidence interval $\delta=10$, and  the real historical data. As one can notice, the resulting point estimates roughly resemble the single run shown by \cite{janssen2009understanding} in Figure 10. In particular, one can observe an initial stationary phase, followed by a steep increase around time $1000$. Then, the number of households stabilizes until time $\sim1120$, when a sharp decrease in population occurs. The number of households recovers around time 1200 and remains constant until time $\sim1260$, when the population rapidly decreases until the final collapse. 
Such a similarity may indicate that the variability under the proposed parameter settings is quite low.

To improve the comparison, in the top-right panel we perform the same estimation exercise fixing, however, the total number of independent replicas that should be performed for each time step to 15. In this way, we are using a sample of independent replication of the same size as the one used by \cite{janssen2009understanding}. As one can notice, a general agreement between the two plots emerges, indicating that, overall, the model presents a quite stable behavior. However, in the time span between 1050 and 1150, the estimates obtained fixing the number of independent replicas to 15 present a confidence interval that is particularly wide. Therefore, in such a time span, these estimates are not very reliable. Indeed, comparing them with those in the top-left panel one notices sensible differences. This may cause erroneous statements in assessing performances. For instance, from the 15-replications exercise one may conclude that in the interval 1050-1150 the model's average number of households is not statistically different from the historical data. This is because the historical data belong to the (too wide) confidence intervals of such time points. But this is an artifact of the low number of independent replicas that the user has set from the outset. Indeed, as shown in the top-left panel, the statement becomes false for refinements of the estimates 
for that period by means of more independent replications. This is made evident in the bottom panel, where we present the number of independent replicas that \mv{} has performed for the transient analysis for each time point. As one can notice, for the 1050-1150 time interval, such a number is between 100 and 600. At the same time, \mv{} efficiently recognizes when a low number of independent replicas is enough, for instance in the first 100 periods it constantly remains below 20 runs. 

In summary, \mv{}'s transient analysis efficiently and reliably estimates the average number of households in each time step, allowing one to avoid erroneous conclusions deriving from statistical hypothesis testing based on an insufficient number of replicas. It also let us better understand the behavior of the model for a given ``best'' choice of parameter values. In what follows, we show that the transient analysis of \mv{} can help in selecting such a ``best'' choice of parameter values. That is, we present  a novel application of the transient analysis: calibrating the model with statistical guarantees. 

\FloatBarrier

\subsection{Model calibration with statistical guarantees}

\begin{table}[!t]
\caption{Values of parameter considered.}
\label{tab:param}
\center
\begin{tabular}{l l}
\toprule
\multicolumn{1}{c}{\emph{Parameter}} & \multicolumn{1}{c}{\emph{Values}} \\
\midrule
Death Age & $\{26,30,34,38\}$\\
End of Fertility Age & $\{26,30,34,38\}$\\
Fission Probability & $\{0.95,0.105,0.115,0.125,0.135,0.145,0.155,0.165,0.175,0.185\}$\\
Harvest Adjustment & $\{0.54,0.56,0.58,0.6\}$ \\
Harvest Variance & $\{0,0.2,0.4,0.6\}$\\
\bottomrule
\end{tabular}
\end{table}

\begin{table}[!h]
\caption{Best combinations of death age (DA), end of fertility age (EFA), harvest adjustment (HA), harvest variance (HV) with indication of the relative estimated loss, width of the confidence interval, estimated variance of the loss, and number of runs used by \mv{} to meet the requirements $\alpha=0.1$ and $\delta=1000$. In red, the best parametrization according to \cite{janssen2009understanding}, while in blue the one with statistical guarantees  obtained using our approach.}
\center
\label{tab:best}
\begin{tabular}{r r r r r r r r r r}
\toprule
\multicolumn{5}{c}{\emph{Parameters}} & \multicolumn{5}{c}{\emph{Obtained measures}}
\\
\cmidrule{1-5} \cmidrule(l){6-10}
\multicolumn{1}{c}{\emph{DA}} & \multicolumn{1}{c}{\emph{EFA }} & \multicolumn{1}{c}{\emph{FP}} & \multicolumn{1}{c}{\emph{HA}} & \multicolumn{1}{c}{\emph{HV}} & \multicolumn{1}{c}{\emph{Estim. Loss}} & \multicolumn{1}{c}{\emph{Width C.I.}} & \multicolumn{1}{c}{\emph{Estim. Var.}} & \multicolumn{1}{c}{\emph{Runs}} & \multicolumn{1}{c}{\emph{p-value}} 
\\
\cmidrule{1-5} \cmidrule(l){6-10}
34 & 30 & 0.185 & 0.56 & 0.4 & 18002 & 783.35 & 1707620 & 32  & 0.7285\\
34 & 34 & 0.145 & 0.56 & 0.4 & 18432 & 973.80 & 12455019 & 144 &  0.1461\\
34 & 34 & 0.165 & 0.56 & 0.4 & 18215 & 847.01 & 1996478 & 32 & 
   0.3376\\
34 & 34 & 0.175 & 0.56 & 0.4 & 17926 & 817.76 & 870417& 16 &
   0.9106  \\
34 & 34 & 0.185 & 0.56 & 0.4 & 18265 & 649.56 & 1174162 & 32  &
   0.2109\\
38 & 30 & 0.185 & 0.56 & 0.4 & 18121 & 807.07 & 2776238 & 48  &
   0.4859\\
\color{red}38 & \color{red}34 & \color{red}0.155 & \color{red}0.56 & \color{red}0.4 & \color{red}17923 & \color{red}817.87 & \color{red}2851035 & \color{red}48 &
 \color{red}0.9191\\
38 & 34 & 0.165 & 0.56 & 0.4 & 18467 & 970.53 & 2621220 &  32  &
   0.1192 \\
38 & 34 & 0.175 & 0.56 & 0.4 & 18335 & 886.03 & 2184662 & 32  &
   0.2023\\
38 & 34 & 0.185 & 0.56 & 0.4 & 18232 & 983.16 & 1258101 & 16  &
   0.3481\\
\color{blue}38 & \color{blue}38 & \color{blue}0.135 & \color{blue}0.56 & \color{blue}0.4 & \color{blue}17889 & \color{blue}771.83 & \color{blue}1657766 & \color{blue}32 &\color{blue}
   1.0000\\
38 & 38 & 0.145 & 0.56 & 0.4 & 18409 & 951.02 & 14553463 & 176 &
   0.1577\\
38 & 38 & 0.155 & 0.56 & 0.4 & 18175 & 940.52 & 2461623 & 32  &
   0.4287\\
38 & 38 & 0.175 & 0.56 & 0.4 & 18373 & 938.87 & 1147316 & 16  &
   0.1768 \\
\bottomrule
\end{tabular}
\end{table}

In the previous application we simply requested \mv{} to estimate the mean and the confidence interval of a given variable of interest at all $t$. However, \mv{} allows the user to estimate mean and confidence interval of a given \emph{function} of the variable of interest recorded at any \emph{interval} of time. 
This can be used to perform a new application of the transient analysis: calibrating the model with statistical guarantees.

More specifically, consider the \emph{loss} of a given combination of parameters $\theta$ computed considering a seed $i$ for the random number generator, it is defined as 
\begin{equation*}
    \mathcal{L}_i(\theta)=\sum\limits_{t=800}^{1350} \left|\hat{h}_{i,t}(\theta)-h_t\right|\,,
\end{equation*}
where $\hat{h}_{i,t}(\theta)$ is the total number of households generated by the model at time $t$ with seed $i$ and $h_t$ is the historically recorded number of households at the same date. Notice that this quantity is actually the same  one that  \cite{axtell2002population} and \cite{janssen2009understanding} use to evaluate a single trajectory under the $L^1$ norm multiplied by a factor of 550 (the total number of steps). 
Our goal is to compute the loss of any $\theta$ trying to reduce as much as possible the intrinsic stochastic variability, that is, the effect of the seeds. In other words, we want to properly estimate such a quantity equipping it with statistical guarantees. To do that, similarly to what done in the previous sub-section for households, we can estimate the loss considering its arithmetic average over a set of $n$ seeds, i.e. $n$ independent replicas, where $n$ is decided in such a way that the estimate belongs to a confidence interval of width $\delta$ with statistical confidence $1-\alpha$. That is, for any $\theta$ considered, we want to compute
\begin{equation*}
    \overline{\mathcal{L}}(\theta;\alpha,\delta)=\sum\limits_{i\in\mathcal{I}(\theta;\alpha,\delta)} \frac{\mathcal{L}_i(\theta)}{|\mathcal{I}(\theta;\alpha,\delta)|}\,,
\end{equation*}
where $\mathcal{I}(\theta;\alpha,\delta)$ is a set of seeds with cardinality $|\mathcal{I}(\theta;\alpha,\delta)|$ such that the confidence interval of $\overline{\mathcal{L}}(\theta;\alpha,\delta)$ with statistical significance $\alpha$ has width at least $\delta$.
Hence, we can use the previously described transient analysis, the only change we have to make is that now \mv{} should evaluate the variable of interest (the loss) only at $t=1350$. This can be easily done by asking \mv{} through its query system to compute the absolute difference between artificial and historical data at any $t$, store its cumulative sum, and statistically evaluate the cumulative sum only at the final time period. In other words, the query language of \mv{} is expressive enough to state those properties. 
Notice that all these steps can be done without changing anything in the model's code, simply querying \mv{}. 

The procedure just described is repeated for any combination of the parameters reported in Table \ref{tab:param}, with the caveat of discarding those combinations where end of fertility age is larger than death age. 
We run the \mv{} transient analysis for all the 1600 parameter combinations setting $\alpha=0.1$ and $\delta=100$. 
Then, the minimization principle used by \cite{axtell2002population} and \cite{janssen2009understanding} would suggest to select the one with lowest loss. However, the estimated values suffer of estimation error, hence, one is not sure whether selecting the one with smallest estimated value is, indeed, the best combination. To account for that, we take the minimum estimated loss as a reference point and discard all the parameter combinations whose estimated loss is significantly larger that the minimum loss according to Welch's t-test \citep{welch1947} for difference of means with $1-\alpha$ confidence level, $90\%$ in our case (we used $\alpha=0.1$). The remaining parameter combinations are those that cannot be considered significantly different from the one with lowest estimated loss and they are shown, together with some technical details provided by \mv{} and the p-value of the Welch's t-test, in Table \ref{tab:best}. As one can notice, the best combination of the parameters highlighted by our exercise is $\{38,38,0.135,0.56,0.4\}$ (in blue in the table). Notably, this does not coincide with the one individuated by \cite{janssen2009understanding} (in red in the table). 
This aligns perfectly with our argument regarding the statistical error associated with estimating loss: given a certain confidence level and estimation precision, there may be parameter combinations whose differences in loss are not statistically significant. Discarding these to focus solely on a single combination may therefore be misleading and result in information loss. For example, identifying patterns within the set of parameter configurations that are not significantly different from the best-performing one can provide insights into the relative importance of different parameters in replicating real data.

Indeed, inspecting the values reported in Table \ref{tab:best} validates the argument made by \cite{janssen2009understanding} regarding the relatively greater importance of harvest-related variables compared to demographic ones in replicating historical population data. All the best parameter combinations include a harvest adjustment of 0.56 and a harvest variance of 0.4, while the other parameters exhibit some variation. This suggests that harvest-related parameters must take specific values for the model to closely match historical data, whereas demographic parameters appear to be less critical.

\begin{figure}[!t]
\centering
\includegraphics[width=0.65\textwidth]{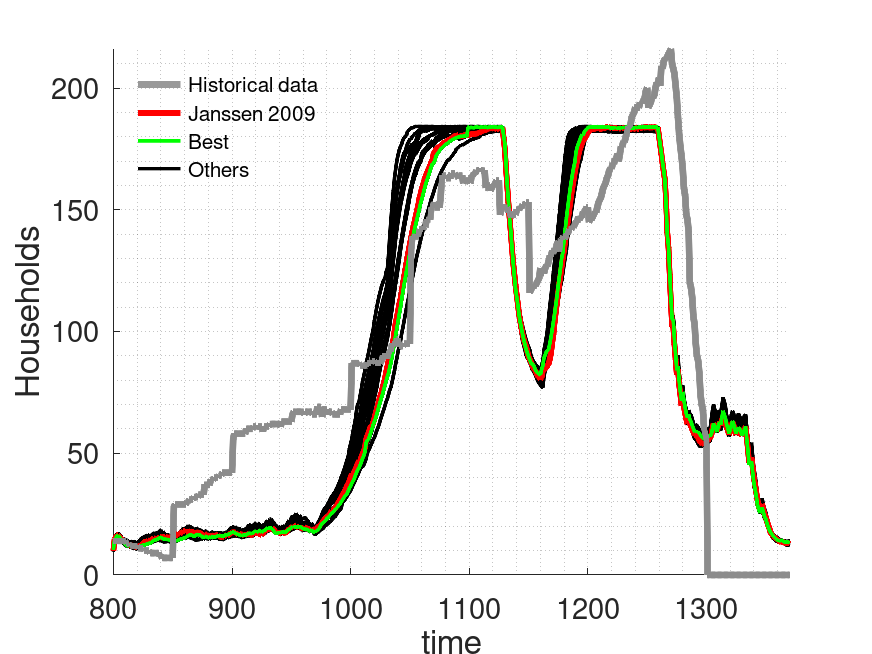}
\caption{Estimated mean number of households at each time step for the 14 best combinations shown in Table \ref{tab:best} together with historical data The trajectory relative to best parameter combination of \cite{janssen2009understanding} is in red, while the one 
according to our estimation is in green.}
\label{fig:anasazi_calib}
\end{figure}

In Figure \ref{fig:anasazi_calib}, we present the time series of the mean number of households resulting from the 14 parameter combinations listed in Table \ref{tab:best}. One can observe that the trajectories are remarkably close at each time step. The largest differences appear between the years 1000 and 1100, a period already identified in previous analyses as one where the model exhibits greater volatility.\footnote{For readers interested in exploring the trajectories in relation to their specific parameter combinations, Figure \ref{fig:anasazi_calib_disag} in Appendix \ref{app:disag} displays the 14 trajectories divided into four panels, with the corresponding parameterizations indicated.}

The calibration procedure with statistical guarantees, as performed here using MultiVeStA's transient analysis, can be refined by reducing $\alpha$ and $\delta$. Doing so narrows the set of best parameter combinations, thereby increasing the precision of the results. However, this refinement also requires greater computational resources, whether in terms of processing power or time. To address this trade-off, we propose a sequential approach: start with relatively large values of $\alpha$ and $\delta$, then iteratively reduce them—applying each refinement to the resulting subset of best combinations—until either a single best combination is identified or the maximum resource allocation is reached. This method allows for a meaningful balance between accuracy and computational cost in the calibration process.

By leveraging \mv{}’s transient analysis, we evaluated a comprehensive set of parameter combinations, explicitly accounting for the uncertainty in loss estimation. Rather than selecting a single ``best'' configuration, our procedure identifies a set of statistically indistinguishable parameter combinations. This is in line with the philosophy of the \emph{Model Confidence Set} approach proposed by \cite{hansen2011model} and recently applied to model calibration by \cite{seri2021model}. Indeed, our method acknowledges that, given estimation uncertainty, multiple specifications may perform equivalently well, and that prematurely narrowing the choice may lead to information loss. This perspective not only highlights the relative importance of certain parameters (such as the dominant role of harvest-related variables in the Artificial Anasazi model) but also enables an iterative and resource-aware refinement of the calibration process. 


\section{Warmup and steady state analysis: Alpha birds model}\label{sec:alphabirds}
We now move our attention towards the steady state analysis capabilities of \mv{}. We aim to achieve statistically reliable results by: (i) identifying the minimal number of independent replications; (ii) determining the duration of the warm-up and of the post-warm-up simulation.

To exemplify \mv{}'s capabilities, we consider the ``Alpha birds'' model, which is an ABM of a territorial bird species whose long-term survival crucially depends on its reproductive characteristics \citep{railsback2012agent}. This model has several advantages. On one hand, it is a typical ABM in ecology, featuring a simplified representation of population dynamics and social behaviors. On the other hand, its steady state features, as will be clarified below, make it an ideal test case for our tools that automatically determine if the steady state has been reached. In addition, this model has already been used by \cite{thiele2014facilitating} to illustrate the features of the RNetLogo package, suggesting that it is a good baseline model that can be approached by various techniques. Interestingly, in the Discussion section of \cite{thiele2014facilitating}, the authors raise the issue that it is often unclear how many independent replications are needed to reliably estimate model properties. They say that 
\[\text{
``\emph{Very likely, just 10 iterations, as used here, will often not be enough.}''
}\]
 As explained in Section \ref{sec:mv}, this is one of the issues addressed by both our transient and steady state analyses.

The model represents a simplified depiction of the dynamics of a territorial bird species that lives in groups and exhibits reproductive suppression. In this system, the alpha couple in each group suppresses the reproduction of subordinate mature birds. A crucial behavior in this context is the decision-making process of subordinate birds regarding when to leave their territory for scouting forays. During these forays, they search for available alpha positions elsewhere. If they fail to find such a position, they return to their home territory. However, scouting forays come with an increased risk of mortality due to raptors.

The model serves as a virtual laboratory for developing a theory on the decision-making process of scouting forays. Different submodels of the foray decision can be implemented, and the resulting outcomes of the full model can be compared to real-world patterns observed in nature. In \cite{railsback2012agent}, the authors utilize patterns generated by a specific version of the model, and the task they propose is to identify the particular submodel they employed. In this article, as in \cite{thiele2014facilitating}, we adopt the simplest version of the model, where the probability that subordinates undertake a scouting foray remains constant.

Thus, a crucial parameter of the model is the scouting probability. There is another important parameter, which is the overall survival probability of any bird. \cite{thiele2014facilitating} explore the response of the model to variations in these two parameters. They focus on three key quantities as key outputs:
\begin{itemize}
    \item \emph{Abundance}, i.e., the total number of birds in all territories;
    \item \emph{Variation}, i.e., the standard deviation of the total number of birds over time;
    \item \emph{Vacancy}, i.e., the fraction of territories with at least one vacant alpha bird.
\end{itemize}

 In their main exercise, the authors explore a range of survival probability between 95\% and 100\%, and a range of scouting probability between 0\% and 50\%.  They consider the mean of the three key outputs across 10 independent replications. Moreover, they run the model for 264 time steps, representing months. These correspond to 22 years. The authors discard the first two years, and then measure Abundance, Variation and Vacancy every 11th month of a year, namely at the 20 time steps 35, 47, 59, ..., 263. 

In the following, we show how \mv{} enables to automatically replicate 
the calibration experiments performed in \cite{thiele2014facilitating}, but we first check if 10 independent replications are indeed sufficient for this analysis, and then if a 2-year warm-up is sufficiently long to wash out the initial warm-up.

\subsection{Number of independent replications}
\label{sec:nruns}
We replicate the same analysis of \cite{thiele2014facilitating} using the manualRD feature of \mv{}, obtaining the values of Abundance, Variation and Vacancy at the same time steps as listed above (35, 47, 59, ..., 263). \mv{} estimations are equipped with confidence intervals. We request that the confidence intervals are bound to a maximum width of: $\delta=5$ for Abundance; $\delta=1$ for Variation; and $\delta=0.04$ for Vacancy. We choose different values of $\delta$ because these three variables vary on three different orders of magnitude. Therefore, if we chose the same value, we would have too much precision for the largest variable and too little precision for the smallest. The results do not depend on the specific values of $\delta$, which are chosen to be around 10\% of the median value of each variable.

As discussed, confidence intervals are computed iteratively by performing blocks of $n$ independent replications. Here we set $n=10$. In this way, the minimum number of replications is 10, as in \cite{thiele2014facilitating}. If 10 replications are not enough for a property, 10 more replications will be performed, and we proceed this way until we identify the minimal amount of independent replications that are needed to obtain statistically valid results with confidence intervals respecting the required $\delta$. 

\begin{figure}[!t]
\centering
\includegraphics[width=1\textwidth]{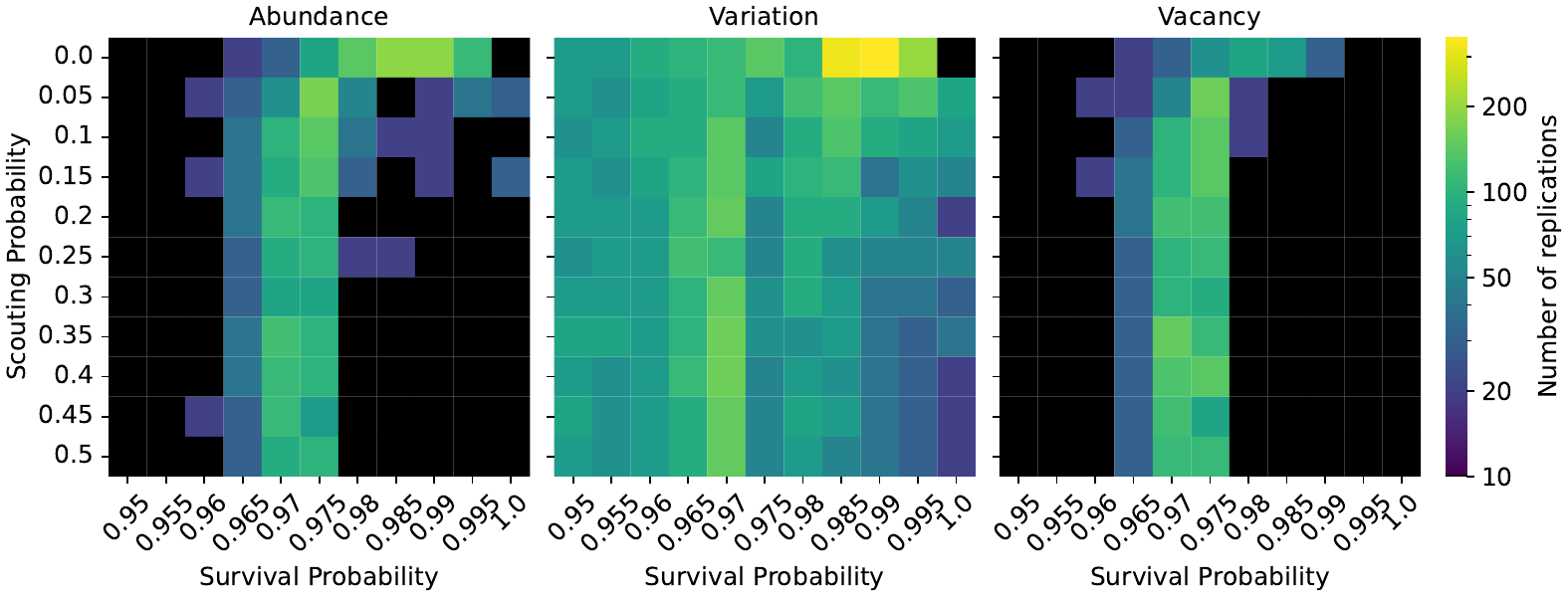}
\caption{Minimal number of independent replications needed to obtain statistically valid results given the required $\delta$. We compute the number of replications for Abundance, Variation and Vacancy, for 121 combinations of the scouting and survival probability parameters. We color in black the parameter combinations for which 10 replications, as used by \cite{thiele2014facilitating}, are sufficient.}
\label{fig:results_alpha_birds_manualRD_nsims}
\end{figure}

Figure \ref{fig:results_alpha_birds_manualRD_nsims} shows how many independent replications are needed for estimating each of Abundance, Variation and Vacancy, for each of the 121 considered parametrization. As we can see, for values of survival probability that lie close to the extremes of the considered interval (around 95\% and 100\%), typically the 10 independent replications considered by \cite{thiele2014facilitating} are sufficient. Instead, for intermediate values of survival probability, namely between 97\% and 97.5\%, the modeler needs to run many more simulations in order to obtain statistically valid results. For instance, for survival probability equal to 97\% and scouting probability equal to 25\%, the analyst needs to run 90 replications to have statistically valid results for Abundance, 100 replications for Variation, and 110 replications for Vacancy. It is also interesting to look at the region of the parameter space  where scouting probability is zero, i.e., birds never venture in scouting forays. In this region, many independent replications are needed for values of survival probability between 97\% and 99\%. For instance, for survival probability equal to 99\%, 350 replications are needed to have statistically valid results for Variation.

This exercise demonstrates that, for many parameter combinations, running 10 simulations for each property for each parametrization as done in \cite{thiele2014facilitating} is not enough to obtain statistically reliable estimations.

\begin{figure}[!t]
\centering
\includegraphics[width=0.9\textwidth]{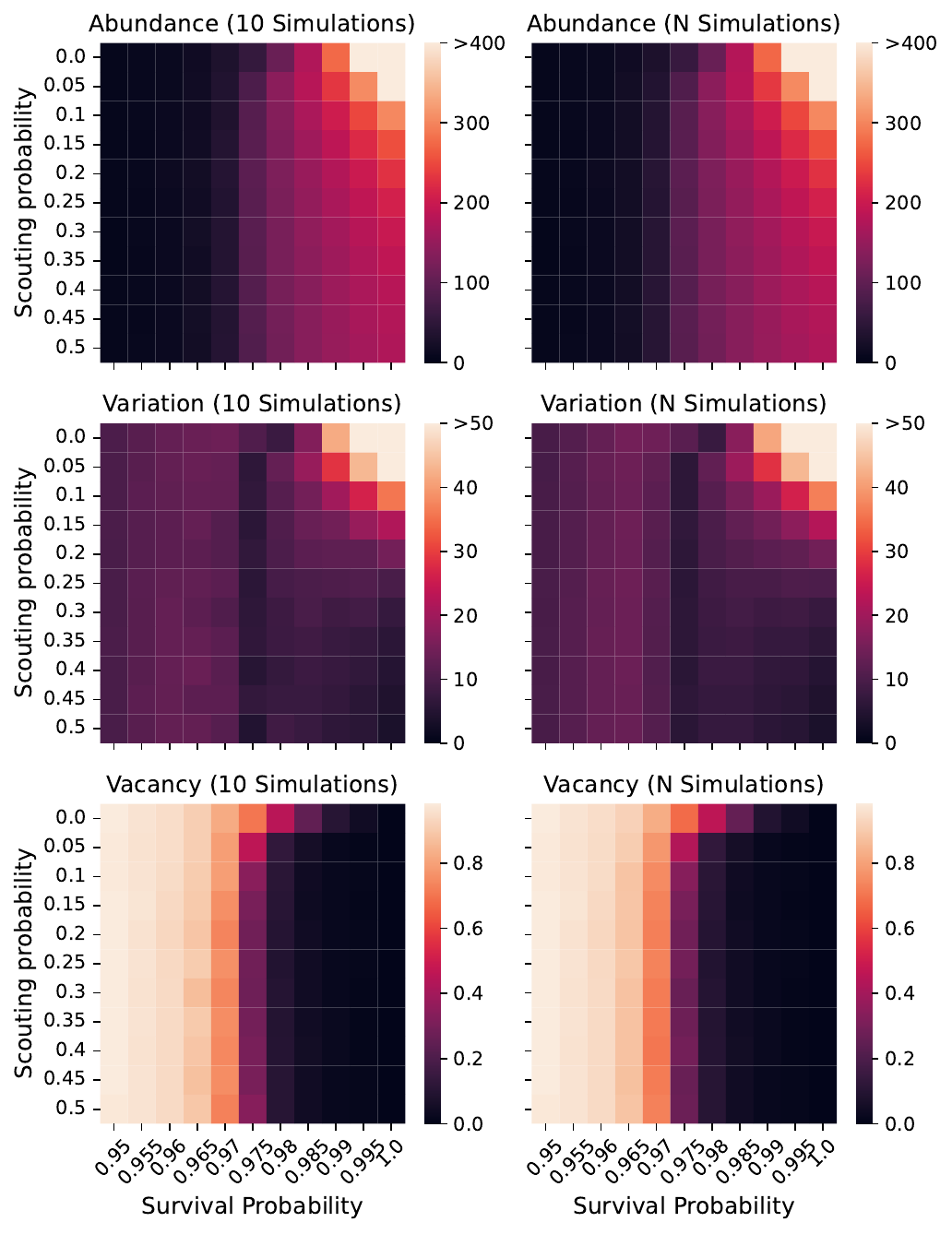}
\caption{Values of Abundance, Variation and Variation for the same parameter combinations shown in Figure \ref{fig:results_alpha_birds_manualRD_nsims}. We show these values both considering 10 simulations only, or letting MultiVesta determine how many independent replications are needed ($N$ independent replications). }
\label{fig:results_alpha_birds_manualRD_scalar}
\end{figure}

To understand why many more independent replications are needed in certain parts of the parameter space, in Figure \ref{fig:results_alpha_birds_manualRD_scalar} we plot the values of Abundance, Variation and Vacancy for all the combinations of parameter values in Figure \ref{fig:results_alpha_birds_manualRD_nsims}. We plot these values both when considering only 10 replications (left), as in \cite{thiele2014facilitating}, and when considering $N$ independent replications (right), as automatically determined by \mv{}. 

We first look at the three variables for 10 replications only (left column). We see that Abundance is always very close to zero for all values of survival probability smaller than 97\%, suggesting that when survival probability is small the bird population tends to turn extinct. Abundance instead reaches higher values, up to 400 birds, for higher survival probability, especially in combination with small scouting probability (recall that scouting forays have higher inherent risk). The situation for Vacancy is specular, as high values for Abundance imply few vacant territories. In the case of Variation, the situation is a bit different, with highest values where Abundance is highest, but lowest values for survival probability equal to 97.5\%, where Abundance and Vacancy substantially change. Overall, these results suggest that the area of the parameter space where most independent replications are needed coincides with the transitions of our key metrics.

It is also interesting to compare the cases of 10 and $N$ independent replications (left to right). We see that there are no major differences between the two cases, but the results of Variation and Vacancy tend to be noisier in the case of 10 independent replications. For instance, for values of survival probability equal to 96.5\% and 97\%, there are sizable fluctuations across values of scouting probability with 10 replications, while the results are much more consistent with $N$ replications. This suggests that, even when results do not qualitatively change, appropriately choosing the number of independent replications makes it possible to estimate the output of the model as a function of the parameters in a smoother way. 

\begin{figure}[!t]
\centering
\includegraphics[width=0.9\textwidth]{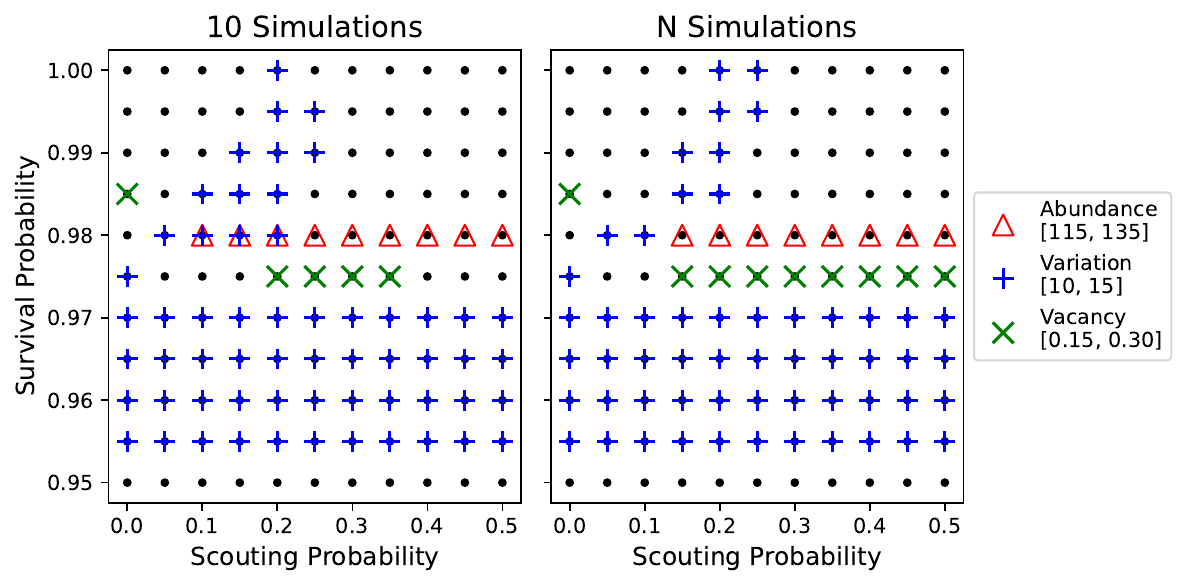}
\caption{Replication of calibration experiment in \cite{thiele2014facilitating}. For each of the 121 parameter combinations, we plot a red triangle when Abundance lies between 115 and 135, a blue plus when Variation is between 10 and 15, and a red cross when Vacancy is between 0.15 and 0.30. The axes are inverted with respect to the previous figures for consistency with Figure 1 in \cite{thiele2014facilitating}. }
\label{fig:results_alpha_birds_manualRD}
\end{figure}

To check the implications of our results, we replicate the first experiment performed in \cite{thiele2014facilitating}. The authors were interested in showcasing some calibration methods and, as a first approach, tested a full factorial design with the 121 parameter combinations that we have been considering. In this design, acceptable values for model outputs were between 115 and 135 for Abundance, 10 and 15 for Variation, and 0.15 and 0.30 for Vacancy. The idea was to check if any parameter combinations would make the model produce an output that was within all three ranges. If taken to characterize a real-world ecosystem, this calibration procedure would find under which parameters the model best describes empirical data. The key result was that no parameter combination fulfilled all three requirements.

The left panel of Figure \ref{fig:results_alpha_birds_manualRD} replicates the calibration results in \cite{thiele2014facilitating}. There are some small differences due to different random seeds, but the overall patterns are very similar. The right panel shows the calibration results when \mv automatically determines the number of independent replications. We see that there are some small differences, for instance green crosses appear for 8 parameter combinations when scouting probability is equal to 97.5\%, whereas green crosses only appear 4 times when running 10 simulations only. Similarly, focusing on 98\% survival probability, under 10 replications there are four values of scouting probability for which variation lies between 10 and 15, while under $N$ replications there are only two such values. Importantly, with $N$ replications it never happens that any two conditions are fulfilled at a time, while this happens for three values of scouting probability when survival probability is 98\%.

In conclusion, we confirm the main result in \cite{thiele2014facilitating}, namely that no parameter combination allow for all three requirements being satisfied at the same time. However, our results showcase that to reach such specific statements it is important to get statistically reliable results, especially in the transition regions of the parameter space.

\subsection{Warmup analysis}
\label{sec:transient}

\begin{figure}[!t]
\centering
\includegraphics[width=1\textwidth]{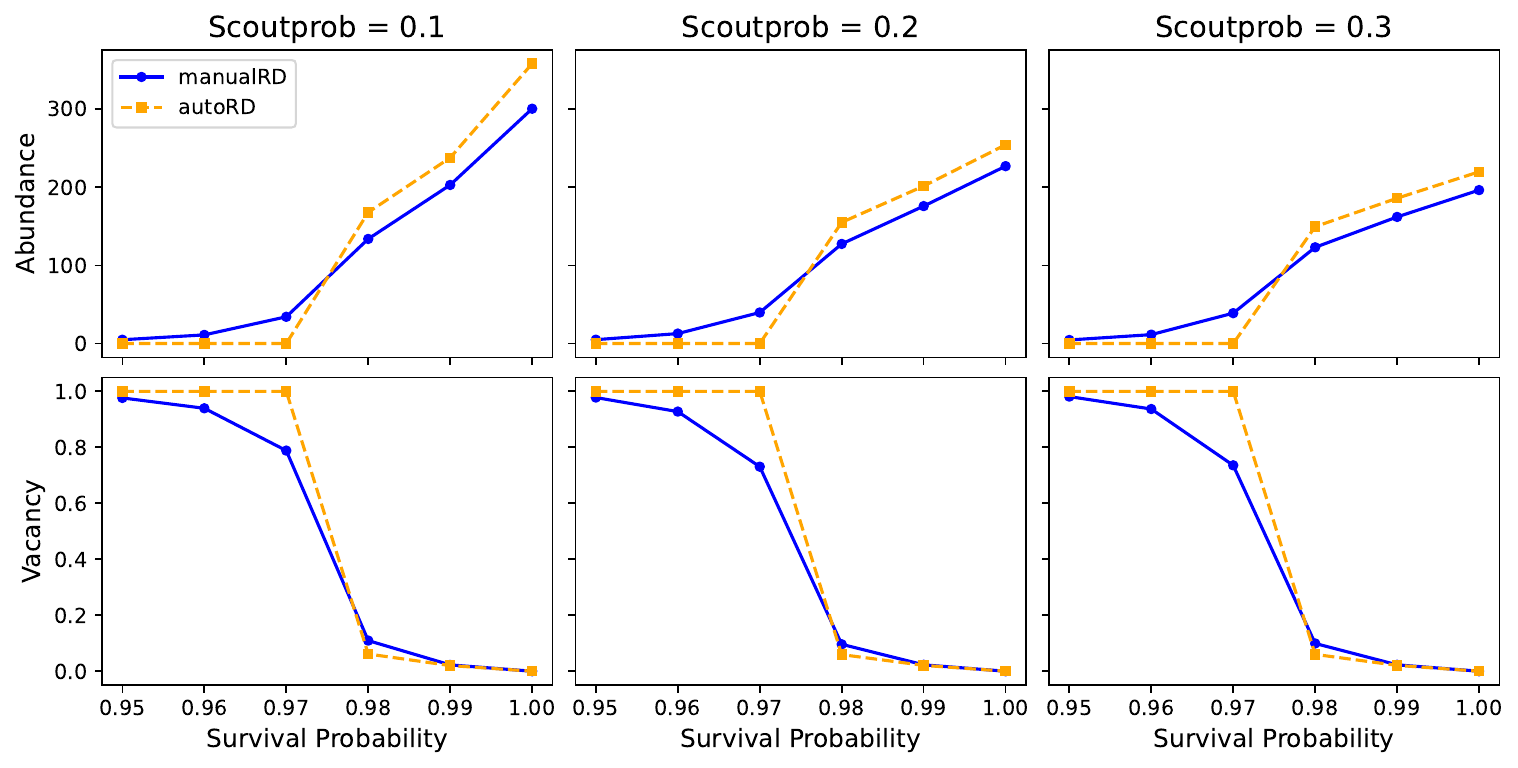}
\caption{Results with a 2-year warm-up vs. an automatically determined warm-up. }
\label{fig:results_manualRD_vs_autoRD}
\end{figure}

As discussed above, \cite{thiele2014facilitating} discard an initial 2-year warm-up. Here, we use \mv{} to test if such a warm-up is sufficiently long for the model to reach the steady state.\footnote{Because it can take very long to achieve statistical guarantees for reaching the steady state, we save simulation time by considering the values of model variables at each time step, instead of the 11-th month of each year.} In case the warm-up was too short, we could check the robustness of the results.

Figure \ref{fig:results_manualRD_vs_autoRD} focuses on Abundance and Vacancy, for all values of survival probability and three selected values of scouting probability.\footnote{These values are the ones for which the algorithm always converges, identifying the warm-up. For some parameter values, \mv{}  cannot detect the reaching of the steady state because it takes very long time to reach a steady state in which the birds species is extinct. Indeed, as birds do not migrate to new territories, when one of the alpha animals in a territory dies, the territory is bound to face local extinction. However, this process may take very long time, and at some point we hit memory limits.} In each panel, we compare the results from manualRD (also shown in Figure \ref{fig:results_alpha_birds_manualRD_scalar}) with the corresponding results from autoRD, which ensure that the steady state has been reached.

Our key result is that, for autoRD there is a sharp transition in Abundance and Vacancy at survival probability equal to 0.97. When survival probability is smaller than 0.97, they take value 0 and 1, respectively, but when survival probability is larger than 0.97, Abundance takes positive value and Vacancy is close to 0. By contrast, for manualRD the transition is much smoother.

This suggests that the smooth transitions that can also be seen in Figure \ref{fig:results_alpha_birds_manualRD_scalar} are really driven by warm-ups. However, in the long run, either the bird species survives, or it collapses.


\FloatBarrier

\section{Discussion}
\label{sec:discussion}

This paper contributes to the ABM literature and community in two complementary ways. On the one hand, it advances methodology by integrating \mv, a statistical model checking framework, with \nl,  one of the most widely used ABM platform. On the other hand, through concrete applications, it demonstrates how this integration enables more reliable and nuanced analyses of ABM outputs.

From a methodological standpoint, the integration provides \nl users with automated statistical tools that can determine the number of simulations, the length of warm-up periods, and the appropriate duration of steady-state runs with statistical guarantees. In so doing, we aim at advancing along the path traced by \cite{thiele2014facilitating,JSSv058i02} and  \cite{mvjedc2022} (among others) and addressing a long-standing limitation in the agent-based modeling community: the difficulty of moving beyond qualitative or anecdotal insights toward reproducible, quantitative, and statistically robust evaluations of model behavior.
Indeed, our integration  represents a significant improvement over the ad hoc rules of thumb typically employed, lowering the barrier to conducting rigorous and reproducible simulation studies. Although the integration requires some familiarity with property specification and statistical settings, it substantially reduces the effort previously needed to implement statistically sound analyses, freeing modelers to focus on substantive modeling questions.

To showcase the potentialities of the integration, we propose two applications, one that relies upon \mv{}'s transient analysis and one that leverages \mv{}'s steady state analysis.
These applications highlight how the resources provided by \mv{} are able to enhance the reliability and interpretability of \nl-based agent-based models. 

By applying the transient analysis techniques to the calibration of the Artificial Anasazi model, we demonstrated that widely used calibration strategies based on a small fixed number of replications may lead to inaccurate conclusions. \mv’s adaptive allocation of simulations provided a richer picture, showing that several parameter combinations can be statistically indistinguishable in their ability to reproduce historical data. This finding reinforces the importance of acknowledging uncertainty in calibration exercises and cautions against the search for a single ``best'' parameter set. The method also confirmed the central role of harvest-related parameters in shaping population dynamics, thereby giving stronger statistical support to earlier (more qualitative) observations.

In the case of the Alpha Birds model, \mv proved effective in detecting inadequacies in previous analysis practices. The assumption of a fixed two-year warm-up period used in earlier studies was shown to be insufficient to capture critical transitions in population dynamics. Our results, instead, revealed sharp shifts in the model behavior near parameter thresholds, which would have been obscured without statistically guided warm-up and steady-state analysis. Moreover, the finding that the number of required replications varies dramatically across regions of the parameter space underscores the risks of adopting uniform rules of thumb for simulation design.

At the same time, some limitations remain. The computational burden of large-scale analyses is non-trivial, and although \mv supports parallelization, scaling to highly complex or computationally intensive models may pose challenges. Additionally, while the integration is seamless at the technical level, non-expert users may still face a learning curve in specifying properties and interpreting statistical outputs. Broader adoption will likely depend on developing more intuitive interfaces and community best practices.

Future research should therefore proceed along two main directions. On the methodological side, an important challenge is to extend the framework to cases of non-ergodicity, where models do not converge to a single stationary distribution and standard steady-state assumptions fail. Many social and economic ABMs exhibit path dependence, multiple equilibria, or lock-in effects, and developing tools to statistically analyze such dynamics would greatly expand the scope of rigorous ABM evaluation. Another possible methodological advancement may regard the investigation of recent statistical model checking techniques that \emph{go beyond means}~\cite{carlos25} by allowing to estimate higher-order moments, quantiles and more.  On the application side, efforts should focus on the efficient analysis of large-scale models, where the number of agents, parameters, or required simulations is substantial. Techniques such as adaptive sampling and surrogate modeling could help balance statistical reliability with computational feasibility, making it possible to apply our approach to models used for real-world policy or large-scale ecological management. Another important avenue concerns advancing calibration and validation practices. While this paper has shown that leveraging \mv's transient analysis one can identify sets of statistically indistinguishable parameter combinations, future work should build systematic workflows for calibrating models against empirical data and validating their predictions under uncertainty. 

In conclusion, the integration of \mv{} with \nl provides both a methodological advance and a practical tool for ABM research. The case studies demonstrate that rigorous statistical analysis is not a mere technical enhancement. 
Indeed, taken together, the two applications show how, embedding statistical rigor in the analysis of ABMs, one  uncovers insights that may otherwise remain hidden, such as the non-uniqueness of calibrated solutions or the presence of discontinuities in model dynamics. By ensuring that conclusions are based on statistically sound evidence and by making statistical rigor accessible within a widely used modeling environment, our integration aims at making ABM studies more transparent, reproducible, and impactful.

\bibliographystyle{jasss}
\bibliography{bibfile} 

\newpage
\appendix

\section{Best parameter combinations of the Artificial Anasazi model, disaggregation of estimated trajectories}
\label{app:disag}
\begin{figure}[!h]
\centering
\includegraphics[width=0.5\textwidth]{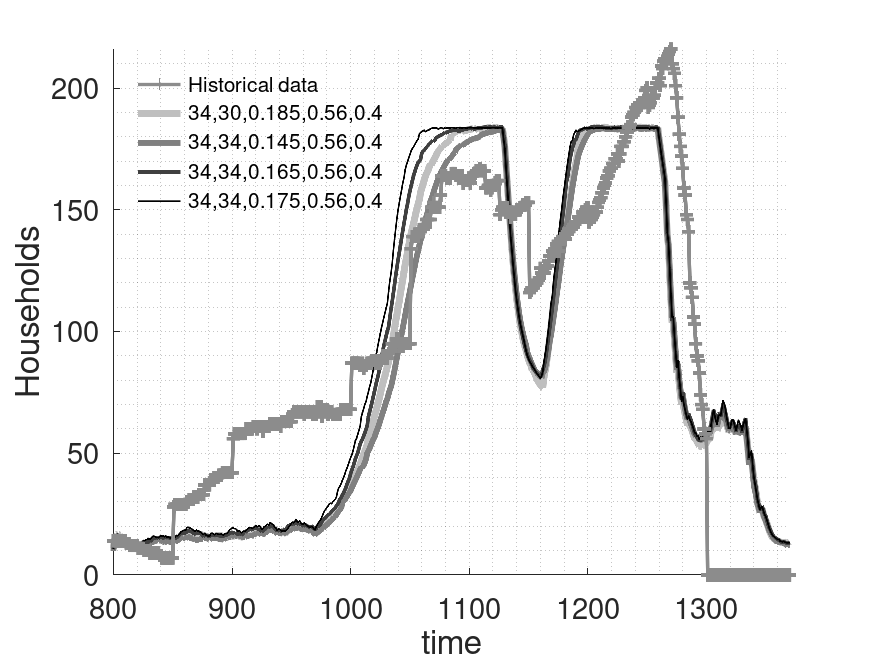}\includegraphics[width=0.5\textwidth]{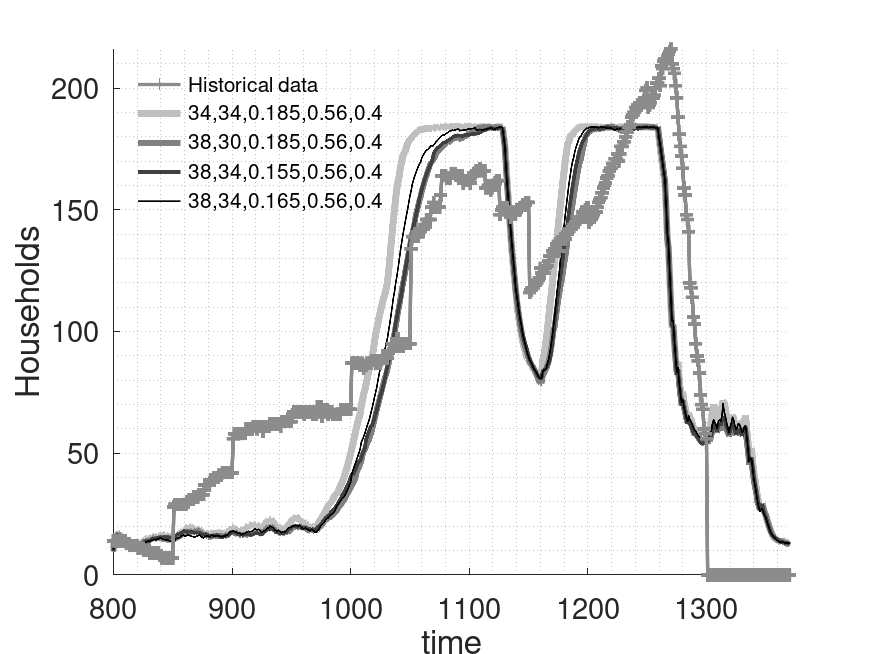}\\\includegraphics[width=0.5\textwidth]{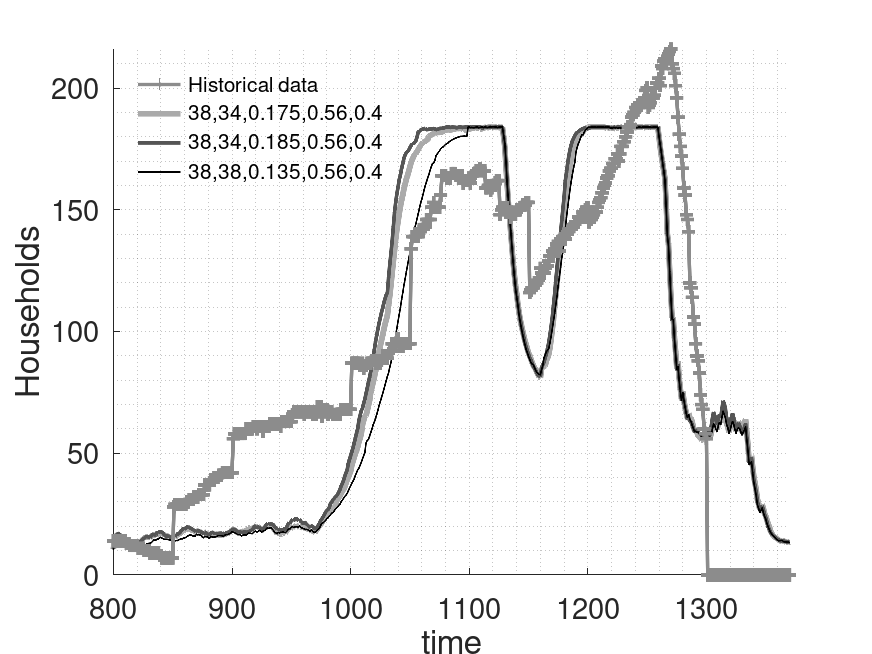}\includegraphics[width=0.5\textwidth]{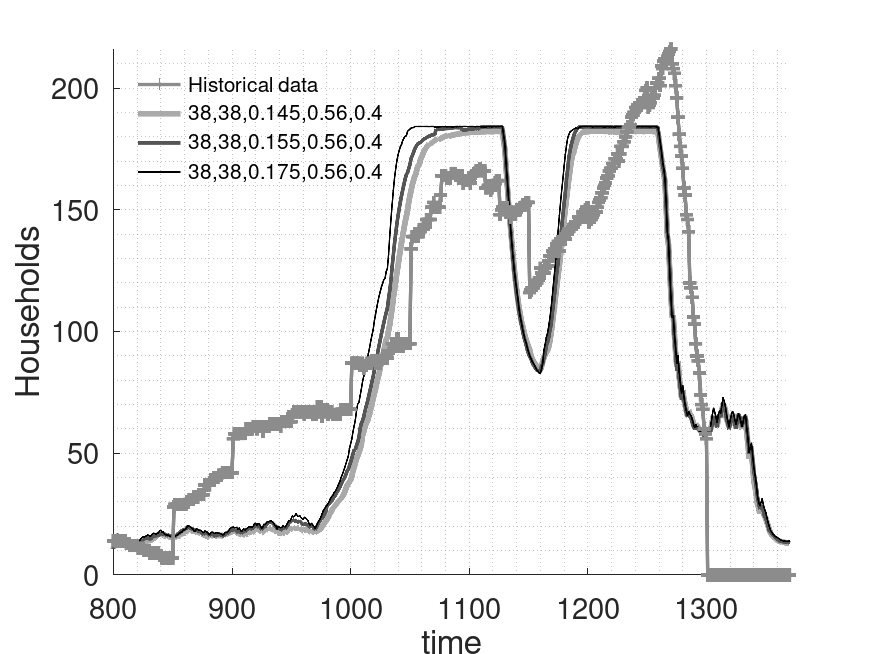}
\caption{Estimated trajectories of the 14 best combinations of parameters.}
\label{fig:anasazi_calib_disag}
\end{figure}

In Figure \ref{fig:anasazi_calib_disag}, we divide into four plots the estimated trajectories of the best combinations according to the procedure previously described.  

\end{document}